\documentclass[aps,prl,twocolumn,superscriptaddress,amsmath,amssymb,letterpaper]{revtex4-1}

\ifx\pdfoutput\@undefined\usepackage[usenames,dvips]{color}
\else\usepackage[usenames,dvipsnames]{color}
\usepackage{graphicx}
\usepackage{bm}
\usepackage{amssymb}
\usepackage{hyperref}
\usepackage{color}

\usepackage{amsmath}

\usepackage[T1]{fontenc}
\usepackage[latin1]{inputenc}

\newcommand{\be}{\begin{equation}}
\newcommand{\ee}{\end{equation}}

\begin{document}

\title{Optical Momentum, Spin, and Angular Momentum in Dispersive Media}

\author{Konstantin Y. Bliokh}
\affiliation{CEMS, RIKEN, Wako-shi, Saitama 351-0198, Japan}
\affiliation{Nonlinear Physics Centre, RSPE, The Australian National University,
Canberra, ACT 0200, Australia}

\author{Aleksandr Y. Bekshaev}
\affiliation{CEMS, RIKEN, Wako-shi, Saitama 351-0198, Japan}
\affiliation{I. I. Mechnikov National University, Dvorianska 2, Odessa, 65082, Ukraine}

\author{Franco Nori}
\affiliation{CEMS, RIKEN, Wako-shi, Saitama 351-0198, Japan}
\affiliation{Physics Department, University of Michigan, Ann Arbor, Michigan 48109-1040, USA}

\begin{abstract}
We examine the momentum, spin, and orbital angular momentum of structured monochromatic optical fields in dispersive inhomogeneous isotropic media. There are two bifurcations in this general problem: the Abraham--Minkowski dilemma and the kinetic (Poynting-like) versus canonical (spin-orbital) pictures. We show that the kinetic Abraham momentum describes the energy flux and group velocity of the wave in the medium. At the same time, we introduce novel canonical Minkowsky-type momentum, spin, and orbital angular momentum densities of the field. These quantities exhibit fairly natural forms, analogous to the Brillouin energy density, as well as multiple advantages as compared with previously considered formalisms. We apply this general theory to inhomogeneous surface plasmon-polariton (SPP) waves at a metal-vacuum interface and show that SPPs carry a ``super-momentum'', proportional to the wave vector $k_{p} > \omega/c$, and a transverse spin, which can change its sign depending on the frequency $\omega$.
\end{abstract}

\vspace*{-0.1cm}

\maketitle

\textit{Introduction.---}
Energy, momentum, and angular momentum (AM) are the main dynamical characteristics of particles and fields, matter and light. These quantities are crucial for understanding the physical properties of objects and predicting their behavior. Electromagnetic waves propagating in optical media are {\it mixed} light-matter (photon-electron) excitations with nontrivial dispersion and dynamical properties. While optical energy density in dispersive isotropic media is described by the well-known Brillouin formula \cite{Jackson}, the characterization of the optical momentum and AM in media is a challenging problem, with the longstanding {\it Abraham--Minkowski} debate lying at its heart \cite{Brevik1979,Pfeifer2007,BarnettLoudon2010,Milonni2010,Kemp2011}.

For pure {\it free-space} light, the Abraham and Minkowski approaches coincide, resulting in the well-known Poynting picture of the momentum and AM of light \cite{Jackson}.
However, even this established formalism produces fundamental difficulties when applied to {\it structured} (inhomogeneous) optical fields. First, the ratio of the Poynting momentum density to the energy density of the field cannot exceed $c^{-1}$ in magnitude, which corresponds to the momentum not exceeding $\hbar \omega/c \equiv \hbar k_0$ per photon. However, in evanescent fields or near optical vortices, the local wave vector (phase gradient) and the corresponding momentum density can be $\hbar k_{\rm loc} > \hbar k_0$, and such ``{\it super-momentum}'' is observed experimentally via light-matter interactions \cite{Huard1978,Matsudo1998,Bliokh2013,Barnett2013}.
Second, the Poynting formalism does not describe separately the {\it spin and orbital} AM of light. At the same time, the spin and orbital degrees of freedom are separately observable in many experiments, and these play important roles in light-matter interactions \cite{Enk1994,ONeil2002, Garces2003,Bliokh2010,OAM,AML,Bliokh2015PR,Aiello2015,Bliokh2015NP}.

The above difficulties with the Poynting free-space formalism can be resolved using the {\it canonical} approach originating from relativistic field theory \cite{Soper,Leader2014}, which explicitly describes the spin and orbital momentum and AM densities for monochromatic free-space fields \cite{Bliokh2015PR,Aiello2015,Berry2009,Barnett2010,Bliokh2013dual,Bliokh2014NC,Bliokh2014NJP}.
This approach deals with the {\it canonical (orbital) momentum density} corresponding to the actual local wave vector of the field, $\hbar k_{\rm loc}$, and the {\it canonical spin AM density} characterizing rotations of the electric and magnetic fields in 3D space. Most importantly, both of these canonical quantities are directly observable in local light-matter interactions via the optical scattering force and torque on small dipole particles or atoms \cite{ONeil2002,Garces2003,Bliokh2015PR,Bliokh2014NC,Bliokh2014PRL,Antognozzi2016,Leader2016}.

Thus, for waves in optical media one can use either the Abraham or Minkowski approaches combined with either the kinetic (Poynting-like) or canonical (spin-orbital) formalisms, i.e., four combinations in total. Here we argue that two of these perfectly describe the kinematic and dynamical properties of optical fields. First, the well-known kinetic Abraham-Poynting momentum density describes the {\it energy flux} and the {\it group velocity} of the wave in the medium. Second, we introduce the novel {\it canonical Minkowski-type momentum, spin, and orbital AM densities}, valid for structured optical fields in dispersive inhomogeneous media. [Note that although the (dispersion-modified) Minkowski momentum is often associated with the canonical one \cite{Dewar1977,Nelson1991,Garrison2004,Stallinga2006,Barnett2010PRL,Dodin2012}, this is the case only for {\it plane waves}, while for generic {\it structured} fields the Poynting vector cannot describe the local wave-vector properties even in free space.]

The novel canonical momentum and spin densities have very natural forms, analogous to the Brillouin energy density and involving the corresponding quantum operators. 
For plane waves in transparent media, they produce natural results associated with the wave vector and polarization helicity. This is consistent with the dispersion-modified Minkowski approach \cite{Dewar1977,Nelson1991,Garrison2004,Stallinga2006,Barnett2010PRL,Dodin2012,
Philbin2011,Philbin2012} and the corresponding plane-wave momentum experiments \cite{Jones1978,Campbell2005}. However, in contrast to previous formalisms, our canonical densities efficiently describe the wave-vector and field-rotation properties of {\it arbitary structured fields}. 

To illustrate this, we apply our general theory to surface plasmon-polaritons (SPPs) (structured fields) at a metal-vacuum interface (an inhomogeneous dispersive medium) \cite{Zayats}. This example has never been considered in the context of the Abraham-Minkowski dilemma. While the kinetic-Abraham-Poynting momentum describes the subluminal group velocity of SPPs \cite{Nkoma}, $v_g<c$, we show that the canonical momentum yields a ``super-momentum'' $\hbar k_{p} > \hbar k_0$ per polariton, both locally and when integrated over the localized field. Strikingly, none of the previous approaches produces this natural result. Moreover, we calculate the canonical {\it transverse spin AM} carried by the SPP. This interesting quantity was introduced earlier using the Abraham-type energy-flux approach \cite{Bliokh2012PRA}, and currently the transverse spin in structured fields is attracting considerable attention \cite{Bliokh2015PR,Aiello2015,Bliokh2015NP,Bliokh2014NC}, promising exciting applications in nano-photonics and quantum optics \cite{Paco2013,Arno2014,Neugebauer2015,Kobus2015,Bliokh2015Science,Lodahl2016}. Remarkably, only now our canonical formalism enables one to calculate the transverse spin of a SPP properly, including the dispersion effects in the metal. We find that the integral transverse spin {\it can change its sign} or vanish depending on the SPP frequency, in contrast to previous calculations.

\textit{General theory.---}
%
We consider monochromatic electromagnetic fields, with complex amplitudes ${\bf E}({\bf r})$, ${\bf H}({\bf r})$, and frequency $\omega$, in an isotropic lossless dispersive (and, in general, inhomogeneous) medium characterized by the permittivity $\varepsilon(\omega,{\bf r})$ and permeability $\mu(\omega,{\bf r})$. The cycle-averaged energy density of the field is given by the well-known Brillouin formula \cite{Jackson}:
\begin{equation}
\tilde W = \frac{{g\,\omega }}{2} \left( {\tilde\varepsilon\, {{\left| {\bf{E}} \right|}^2} + \tilde\mu\, {{\left| {\bf{H}} \right|}^2}} \right)\,,
\label{eq:energy}
\end{equation}
where we use Gaussian units with $g=(8\pi\omega)^{-1}$, 
$\tilde\varepsilon = \varepsilon + \omega \,d\varepsilon /d\omega$, 
$\tilde\mu = \mu + \omega \,d\mu /d\omega$, 
and we mark all dispersion-modified quantities by a tilde. 
The kinetic Abraham momentum density of the field is determined by the Poynting vector:
\begin{equation}
{\bm{\mathcal P}}_A = g\,{k_0}\,{{\rm Re}} \left( {\bf E}^*\! \times {\bf H} \right)\,.
\label{eq:Abraham}
\end{equation}
This quantity describes the {\it energy flux} (rather than momentum) density \cite{Soper,Dewar1977,Dodin2012}. The ratio of the integral Abraham momentum (\ref{eq:Abraham}) to the integral energy (\ref{eq:energy}) determines the group velocity of a wave packet in the medium:
\begin{equation}
{{\bf{v}}_g} = \frac{c^2 \left\langle {\bm{\mathcal P}}_A \right\rangle}{{\left\langle {\tilde W} \right\rangle }}\,,
\label{eq:group}
\end{equation}
where $\left\langle ... \right\rangle$ denotes the proper spatial integration of the densities. As we show below, Eq.~(\ref{eq:group}) agrees with the usual $\partial\omega/\partial {\bf k}$ definition even for inhomogeneous waves in inhomogeneous media  \cite{Dewar1977,Dodin2012,Nkoma}.
 
We now put forward the canonical momentum density of the field:
\begin{equation}
\tilde{\bf P} = \frac{g}{2}\,{{\rm Im}} \left[ {\tilde\varepsilon \,{{\bf{E}}^*}\! \cdot \left(\bm{\nabla}\right){\bf{E}} + \tilde\mu \,{{\bf{H}}^*}\! \cdot \left(\bm{\nabla}\right){\bf{H}}} \right]\,,
\label{eq:momentum}
\end{equation}
and also the canonical spin and orbital AM densities:
\begin{equation}
\tilde{\bf S} = \frac{g}{2}\,{{\rm Im}} \left[ {\tilde\varepsilon \,{{\bf{E}}^*}\! \times {\bf{E}} + \tilde\mu \,{{\bf{H}}^*}\! \times {\bf{H}}} \right]\,,~~\tilde{\bf L} = {\bf r} \times \tilde{\bf P}\,.
\label{eq:spin}
\end{equation}
Equations (\ref{eq:momentum}) and (\ref{eq:spin}) are the central expressions of this work, which describe the actual momentum, spin, and orbital AM densities carried by generic {\it structured} optical fields in an {\it dispersive inhomogeneous} medium. Below we consider several remarkable properties and applications of these quantities. 

1. In the vacuum, $\tilde\varepsilon = \tilde\mu = 1$, and the densities (\ref{eq:momentum}) and (\ref{eq:spin}) coincide with the corresponding canonical densities for free-space fields \cite{Bliokh2015PR,Aiello2015,Berry2009,Barnett2010,Bliokh2013dual,Bliokh2014NC,
Bliokh2014NJP,Bliokh2014PRL}, which are consistent with directly-observable properties of structured optical fields \cite{Huard1978,Matsudo1998,Bliokh2013,Barnett2013,ONeil2002,Garces2003,Antognozzi2016,
Leader2016}.

2. Equations (\ref{eq:momentum}) and (\ref{eq:spin}) exhibit a pleasing similarity with the Brilluoin energy density (\ref{eq:energy}). Together, these can be written as a consistent set of dynamical quantities using the corresponding quantum-mechanical operators:
\begin{equation}
\tilde{W} = \psi^{\dag}\!\left(\omega\right)\psi\,,~~\tilde{\bf P} = {\rm Re}\!\left[ \psi^{\dag}\!\left(\hat{\bf p}\right)\psi\right]\,,~~\tilde{\bf S} = \psi^{\dag}\!\left(\hat{\bf S}\right)\psi\,.
\label{eq:operators}
\end{equation}
Here $\hat{\bf p}\! =\! -i\, \bm{\nabla}$ and $\hat{\bf S}$ are the momentum and spin-1 operators \cite{Bliokh2015PR,Berry2009,Bliokh2013dual,Bliokh2014PRL}, whereas the 6-component ``wavefunction'' is $\psi  = \sqrt{g/2}\, {\left( {\sqrt {\tilde \varepsilon } \,{\bf{E}},\sqrt {\tilde \mu } \,{\bf{H}}} \right)^T}$. Notably, this quantum-like formalism exactly coincides with the one recently introduced by Silveirinha for calculations of other electromagnetic bi-linear forms in dispersive media  \cite{Silveirinha2015,Silveirinha2017}.

3. Consider the simplest case of an electromagnetic plane wave in a homogeneous transparent medium. Using the Maxwell equations, we readily obtain the ratios of the densities (\ref{eq:Abraham}), (\ref{eq:momentum}), and (\ref{eq:spin}) to the energy density (\ref{eq:energy}):
\begin{equation}
\frac{{\bm{\mathcal P}}_A }{\tilde{W}} = \frac{1}{n_p n_g}\frac{\bf k}{\omega}\,,~~
\frac{\tilde{\bf P}}{\tilde{W}} = \frac{\bf k}{\omega}\,,~~\frac{\tilde{\bf S}}{\tilde{W}} = \frac{\sigma}{\omega}\frac{\bf k}{k}\,,
\label{eq:planewave}
\end{equation}
where ${\bf k}$ is the wave vector in the medium, $n_p = \sqrt{\varepsilon\mu}$ 
and $n_g = n_p + \omega\, d n_p / d \omega$ are the phase and group refractive indices of the medium, and $\sigma$ is the polarization helicity (the third Stokes parameter). Using $k=n_p k_0$, we see that the first Eq.~(\ref{eq:planewave}) provides the local counterpart of the group-velocity Eq.~(\ref{eq:group}), yielding $v_g = c/n_g < c$. Assuming the quantization of energy as $\hbar\omega$ per photon, the second Eq.~(\ref{eq:planewave}) yields the canonical momentum corresponding to $\hbar {\bf k}$ per photon. This natural result exactly coincides with that obtained from the {\it dispersion-modified Minkowski momentum}, sometimes using rather cumbersome expressions \cite{Nelson1991,Garrison2004,Stallinga2006,Barnett2010PRL,Dodin2012,Philbin2011}.
Finally, the spin AM density in Eq.~(\ref{eq:planewave}) also acquires very the natural form $\hbar \sigma {\bf k}/k$, exactly as one would expect for a photon. To the best of our knowledge, this simple result was previously derived for dispersive media only in \cite{Philbin2012}, using rather nontrivial calculations.

4. Previously, dispersion-modified Mikowski-type momentum and AM densities,  ${\tilde{\bm{\mathcal P}}_M }$ and ${\tilde{\bm{\mathcal J}}_M }$, were derived in the most general form, using a Lagrangian-Noether formalism, by Philbin and Allanson \cite{Philbin2011,Philbin2012}. Since these works used the symmetrization of the energy-momentum tensor \cite{Soper,Bliokh2013dual}, they produced {\it kinetic} quantities, yielding the Poynting momentum and {\it total} AM in free space. The explicit lengthy expressions of \cite{Philbin2011,Philbin2012} for monochromatic fields can be written as 
\begin{equation}
{\tilde{\bm{\mathcal P}}_M }\! = \varepsilon\mu {\bm{\mathcal P}}_A\! + \{{\rm disp.}\}\,,~
{\tilde{\bm{\mathcal J}}_M }\! = {\bf r} \times {\tilde{\bm{\mathcal P}}_M }\! + \{{\rm disp.}\}\,,
\label{eq:Philbin}
\end{equation}
where $\{{\rm disp.}\}$ indicates dispersion-related terms. Remarkably, as we show elsewhere \cite{arXiv}, the canonical momentum density (\ref{eq:momentum}) differs from the kinetic one (\ref{eq:Philbin}) by a curl of a vector field ${\bf S}$: $\tilde{\bf P} = {\tilde{\bm{\mathcal P}}_M } + {\bm\nabla} \times {\bf S}/2$, which does not contribute to integral momentum values and conservation laws. Moreover, the total AM (\ref{eq:Philbin}), integrated over volume for a localized field, coincides with the sum of the integral values of the canonical spin and orbital AM (\ref{eq:spin}):
\begin{equation}
\left\langle{\tilde{\bm{\mathcal P}}_M }\right\rangle = \left\langle{\tilde{\bf P}}\right\rangle\,,~
\left\langle{\tilde{\bm{\mathcal J}}_M }\right\rangle =  \left\langle{\tilde{\bf S}}\right\rangle\ + 
\left\langle{\tilde{\bf L}}\right\rangle\,,
\label{eq:integral}
\end{equation}
Thus, our momentum, spin, and orbital AM densities are canonical counterparts of the kinetic Minkowski-type quantities of Philbin and Allanson \cite{Philbin2011,Philbin2012}. The advantages of our quantities are: (i) considerably simpler form, (ii) explicit spin-orbital separation, and (iii) description of canonical properties (e.g., ``super-momentum'') in free-space and media. 

5. Being derived from Neother's theorem \cite{Philbin2011,Philbin2012}, the kinetic quantities (\ref{eq:Philbin}) are {\it conserved} in media with the corresponding translational/rotational symmetries. Due to Eqs.~(\ref{eq:integral}), this is also true for our canonical quantities (\ref{eq:momentum}) and (\ref{eq:spin}). This can also be seen from the plane-wave Eqs.~(\ref{eq:planewave}), valid for paraxial optical beams. Indeed, the momentum $\hbar {\bf k}$ per photon and spin $\hbar \sigma {\bf k}/k$ underpin the momentum and AM conservation laws for the optical beam reflection/refraction at a planar interface between two media \cite{Onoda2004,Bliokh2006,Bliokh2013JO} (the simplest example being Snell's law \cite{Jackson}).

6. Most importantly, considering an example of surface plasmon-polaritions at a metal-vacuum interface, elswhere we show that both kinetic and canonical momentum and AM densities (\ref{eq:momentum}), (\ref{eq:spin}), and (\ref{eq:Philbin}) can be {\it derived microscopically}  \cite{arXiv}. This allows one to separate the microscopic electromagnetic-field and electron-matter contributions. In particular, it is the electron contributions that are responsible for the dispersion-related terms. 

7. Moreover, considering a non-magnetic medium, $\mu=1$, the dispersion corrections in the spin AM density (\ref{eq:spin}) produces the {\it magnetization} in the medium due to the inverse Faraday effect \cite{Jackson,Hertel}: ${\bf M}\propto \omega\, d\varepsilon/d\omega\, {\rm Im}\!\left( {\bf E}^*\! \times {\bf E} \right)$. In turn, this magnetization generates the {\it direct magnetization current} ${\bf j}_{\rm magn}\! = c\, {\bm \nabla} \times {\bf M}$. Remarkably, the momentum density carried by the electrons in this direct current exactly corresponds to the difference between the kinetic Abraham and Minkowski-type momenta \cite{arXiv}: ${\bm{\mathcal P}}_{\rm magn} = (m/e)\,{\bf j}_{\rm magn} = {\bm{\mathcal P}}_A - \tilde{\bm{\mathcal P}}_M$, where $m$ and $e<0$ are the electron mass and charge.

8. Note that the canonical momentum and spin densities (\ref{eq:momentum}) and (\ref{eq:spin}) have the {\it dual-symmetric} form, keeping the electric and magnetic contributions on an equal footing. In free space, such formalism was recently suggested and extensively discussed 
in \cite{Bliokh2015PR,Aiello2015,Berry2009,Barnett2010,Bliokh2013dual,Bliokh2014NC,
Bliokh2014NJP,Bliokh2014PRL,Cameron2012,Cameron2012II}, mostly from aestetic reasons (``electric-magnetic democracy'' \cite{Berry2009}) rather than real physical arguments.
Alternatively, in free space, one can use the {\it electric-biased} formalism, originating from the dual-asymmetric form of the standard electromagnetic field Lagrangian \cite{Soper,Leader2014,Bliokh2013dual,Bliokh2014NJP,Antognozzi2016,Leader2016}. It yields ``{\it double-electric}'' momentum and spin densities ${\bf P}^\prime = g\, {\rm Im}\!\left[ {\bf E}^*\!\cdot ({\bf\nabla}) {\bf E} \right]$ and ${\bf S}^\prime = g\, {\rm Im}\!\left( {\bf E}^*\!\times {\bf E} \right)$. The choice of the electric-biased quantities does not affect the free-space integral momentum and spin values \cite{Barnett2010,Bliokh2013}. However, this is {\it not} the case for fields in dispersive media. First, the integral electric and magnetic contributions to the momentum and spin AM are {\it not} equal to each other anymore. Second, the dispersion-related terms in both canonical and kinetic characteristics (\ref{eq:momentum}), (\ref{eq:spin}), and (\ref{eq:Philbin}) have {\it fixed dual-symmetric form}, confirmed by both the macroscopic Noether-theorem derivation \cite{Philbin2011,Philbin2012} and microscopic calculations \cite{arXiv}. This allows one to discriminate between the dual-symmetric and electric-biased approaches, in favor of the dual-symmetric one.

\begin{figure}
\centering
\includegraphics[width=\columnwidth]{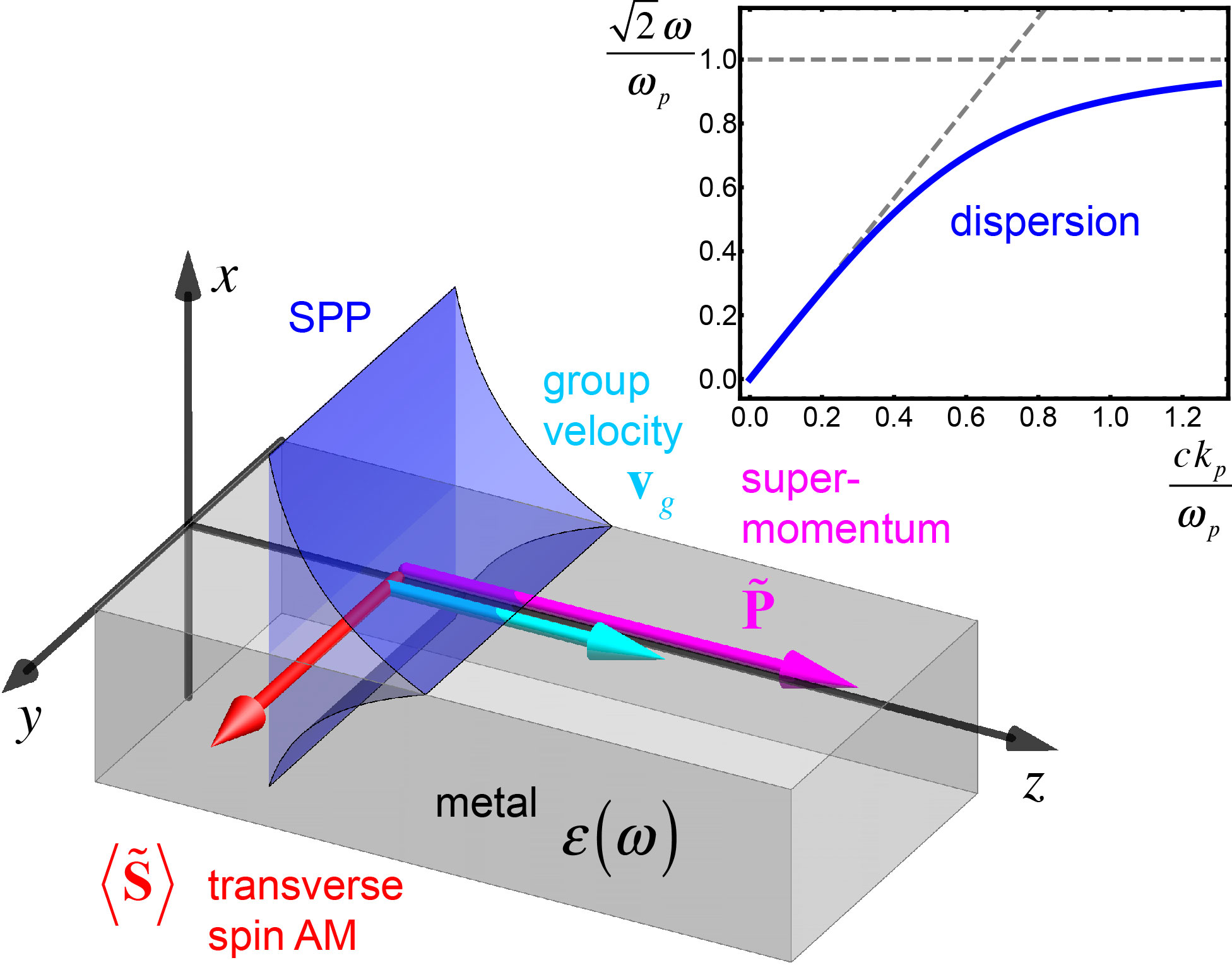}
\caption{Geometry and main properties of a surface plasmon-polariton (SPP) at a metal-vacuum interface.}
\label{fig:SPP}
\end{figure}

\textit{Application to surface plasmon-polaritons.---}
%
We now apply our general theory to an example of a surface plasmon-polariton (SPP) (essentially inhomogeneous field) at a metal-vacuum interface (strongly inhomogeneous dispersive medium) \cite{Zayats}. The geometry of the problem is shown in Fig.~\ref{fig:SPP}, the metal is characterized by the permittivity $\varepsilon (\omega) = 1 - \omega_{p}^{2}/\omega^2$, with $\omega_{p}$ being the electron plasma frequency, whereas $\mu = 1$. 
The SPP is a transverse-magnetic (TM) wave that exists when $\varepsilon<-1$, propagates 
along the interface with the wave vector $k_p = \sqrt{\varepsilon /(1+\varepsilon)}\, k_0 > k_0$, and decays exponentially away from the interface. From here, one can derive the SPP dispersion relation $\omega(k_p)$, shown in Fig.~\ref{fig:SPP}. 

\begin{figure}
\centering
\includegraphics[width=\columnwidth]{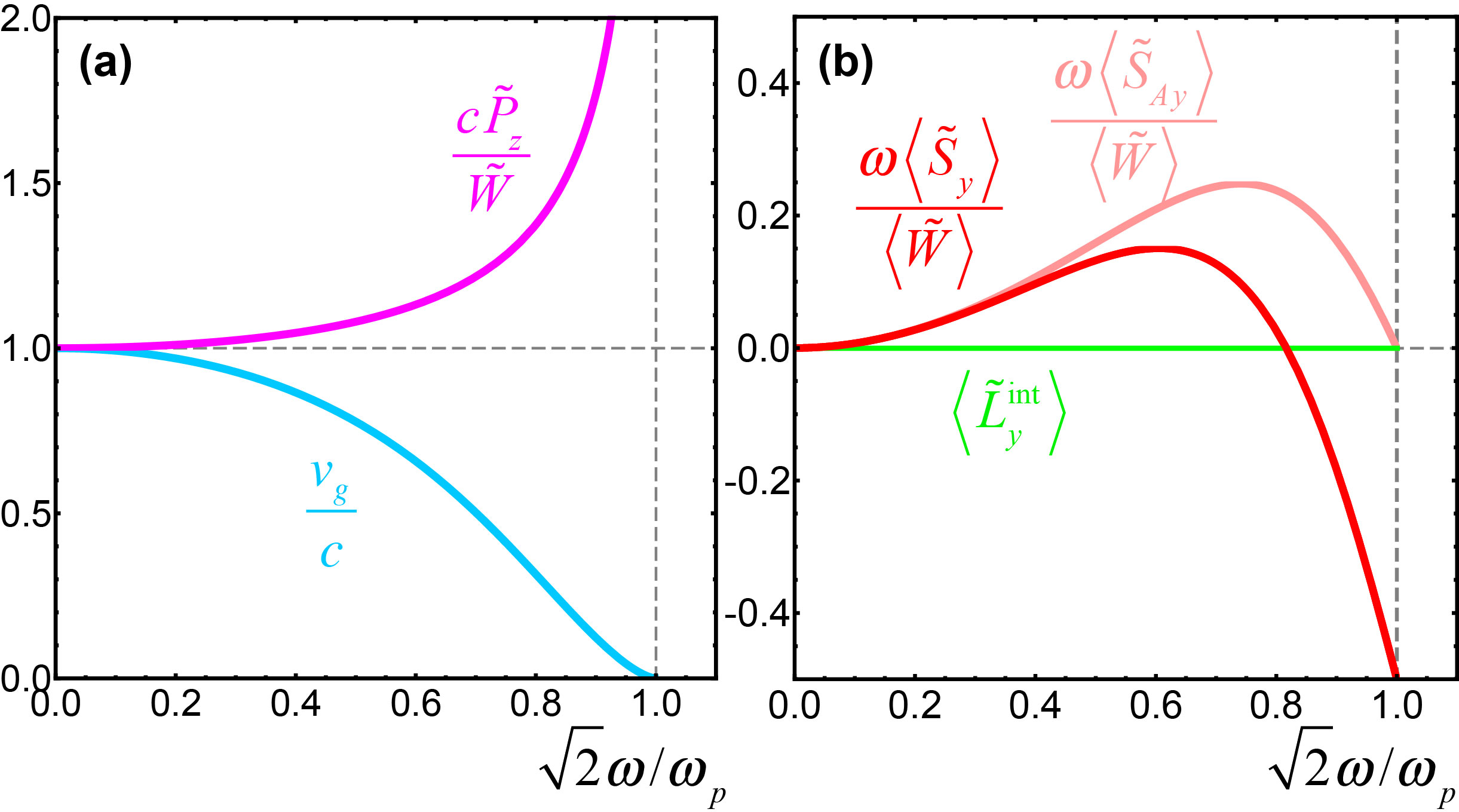}
\caption{(a) The subluminal group velocity (\ref{eq:group}), (\ref{eq:SPPgroup}), and the canonical ``super-momentum'' (\ref{eq:SPPmomentum}) of a SPP versus frequency. (b) The canonical transverse spin of a SPP (\ref{eq:SPPspin}), the previously-calculated Abraham-type spin \cite{Bliokh2012PRA}, and vanishing intrinsic orbital AM of a SPP versus frequency.}
\label{fig:plots}
\end{figure}

Substituting the electric and magnetic fields ${\bf E}$ and ${\bf H}$ of the SPP  \cite{Zayats,Nkoma,Bliokh2012PRA} in Eqs.~(\ref{eq:energy})--(\ref{eq:group}), with $\langle ... \rangle$ denoting the $x$-integration across the interface, we find that the group velocity of SPPs (\ref{eq:group}) exactly coincides with the standard definition \cite{Nkoma}:  
\begin{equation}
{{\bf{v}}_g} = \frac{{\left( {-1 - \varepsilon } \right)^{3/2}\sqrt { - \varepsilon } }}{{1 + {\varepsilon ^2}}}\,c\, \bar{\bf z}
= \frac{\partial \omega}{\partial k_p}\, \bar{\bf z}\,,
\label{eq:SPPgroup}
\end{equation}
where $\bar{\bf x}$, $\bar{\bf y}$, and $\bar{\bf z}$ are the unit vectors of the corresponding coordinates. Naturally, this velocity is always subluminal: $v_g < c$, Fig.~\ref{fig:plots}(a). This confirms the association of the Abraham-Poynting momentum with the group velocity in the general case of inhomogeneous fields and inhomogeneous dispersive media.

Next, we calculate the energy (\ref{eq:energy}) and canonical momentum (\ref{eq:momentum}) of the SPP field. Since all the field components share the same $\exp\!\left( i k_p z \right)$ phase factor, we immediately find, for both the local and integral quantities:
\begin{equation}
\frac{\tilde{\bf P}}{\tilde{W}} = \frac{\left\langle\tilde{\bf P} \right\rangle}{\left\langle\tilde{W}\right\rangle} = \frac{k_p}{\omega}\,\bar{\bf z}\,.
\label{eq:SPPmomentum}
\end{equation}
This equation is analogous to the plane-wave homogeneous-medium Eq.~(\ref{eq:planewave}), but now it is valid for the strongly-inhomogeneous SPP case. This means that the SPP carries ``{\it super-momentum}'' $\hbar k_p > \hbar k_0$ per polariton, both locally and integrally, Fig.~\ref{fig:plots}(a). Surprisingly, none of the previous approaches can provide this seemingly simple result, which is essentially the de Broglie relation! 
In particular, the kinetic Minkowski-type momentum (\ref{eq:Philbin}) agrees with it integrally, Eq.~(\ref{eq:integral}), but it cannot explain the local super-momentum in the evanescent vacuum part of the SPP field, which is observed experimentally \cite{Huard1978,Matsudo1998,Bliokh2013}. Note also that the canonical momentum (\ref{eq:SPPmomentum}) is always directed along the SPP propagation, while the Poynting-Abraham momentum (\ref{eq:Abraham}) is directed backwards inside the metal \cite{Nkoma}.

We finally calculate the {\it transverse spin} AM of the SPP: a quantity which has recently attracted considerable attention \cite{Bliokh2015PR,Aiello2015,Bliokh2015NP,Bliokh2012PRA,Paco2013,Bliokh2015Science,Lodahl2016}, but has never been properly calculated including the dispersion corrections in the metal.
Substituting the SPP fields into Eqs.~(\ref{eq:energy}) and (\ref{eq:spin}), we obtain for the integral spin value:
\begin{equation}
\frac{\left\langle {\tilde{\bf S}} \right\rangle }{\left\langle {\tilde W} \right\rangle } = 
 \frac{{\left( {-2 - \varepsilon } \right)\sqrt { - \varepsilon } }}{{1 + {\varepsilon ^2}}}\,\frac{1}{\omega }\,\bar{\bf y}\,.
\label{eq:SPPspin}
\end{equation}
This equation differs drastically from the plane-wave Eqs.~(\ref{eq:planewave}), showing that now we deal with a {\it structured}-light property, vanishing for a plane TM wave. Equation (\ref{eq:SPPspin}) also differs considerably from the previous calculations of the transverse spin, based on the spin-orbital decomposition of the Abraham-Poynting energy flux \cite{Bliokh2012PRA} and neglecting dispersion effects \cite{Bliokh2015Science}, Fig.~\ref{fig:plots}(b). In particular, the $y$-component of the spin (\ref{eq:SPPspin}) is positive for $\omega < \omega_p/\sqrt{3}$ and {\it negative} for $\omega_p/\sqrt{3} < \omega < \omega_p/\sqrt{2}$. The absolute value of the transverse spin does not exceed $\hbar/2$ per polariton, because it has only the electric-field contribution but not the magnetic one.

Remarkably, the magnetization of the metal, mentioned above, means that a SPP carries not only the spin AM but also a transverse {\it magnetic moment}. Microscopic calculations, presented elswhere \cite{arXiv}, yield the magnetic moment ${\bm \mu} 
\equiv \omega\,\langle {\bf M} \rangle / \langle \tilde{W} \rangle = [2\sqrt{-\varepsilon}/(1+\varepsilon^2)]\, \mu_B\,\bar{\bf y}$ per polariton, where $\mu_B = |e|\hbar / 2mc$ is the Bohr magneton.

One can also calculate the {\it orbital} AM density for the SPP using Eq.~(\ref{eq:spin}). However, this quantity is {\it extrinsic}, i.e., dependent on the choice of the coordinate origin. It makes sense to calculate the {\it intrinsic} part of the integral orbital AM determined with respect to the SPP center of energy $\langle x \rangle$: 
$\left\langle {\tilde{L}}^{\rm int}_y \right\rangle = \int{\left(x-\langle x \rangle\right)\! \tilde{P}_z}\,dx = 0$, Fig.~\ref{fig:plots}(b). It vanishes because of the proportionality (\ref{eq:SPPmomentum}) between the energy and canonical-momentum densities.
Note, however, that the second Eq.~(\ref{eq:integral}) fails in the case of a single SPP wave, and the transverse spin (\ref{eq:SPPspin}) cannot be found via the $x$-integration of the kinetic AM (\ref{eq:Philbin}) taken with respect to $\langle x \rangle$. This is because Eqs.~(\ref{eq:integral}) are valid, rigorously speaking, only for 3D-localized solutions, while a single SPP wave is localized only in one $x$-dimension; considering a $z$-localized SPP wavepacket would fix this discrepancy. This provides one more reason to use the canonical rather than the kinetic picture for the spin and orbital AM calculations.

\textit{Conclusions.---}
We have provided the general theory of the canonical momentum, spin and orbital AM, which is valid for inhomogenous (but monochromatic) optical fields in dispersive and inhomegenous (but isotropic and lossless) media. Our approach combines mathematical simplicity with physical generality, and none of the previously-used definitions of the momentum and AM densities is able to reproduce all the seemingly simple results obtained in this work. The remarkable features and advantages of the suggested formalism are explained in points 1--8 in the main text. 
We have considered surface plasmon-polaritons only as the simplest example of the application of our theory, where other approaches fail. Taking into account both material and structured-light properties is crucial in a variety of nano-optical and photonic systems, including photonic crystals, metamaterials, and optomechanical systems. Our theory provides an efficient toolbox for the description of dynamical properties of light in such systems.
We hope that this formalism will become as useful as the Brillouin energy-density expression.


\begin{acknowledgments}
This work was supported by the RIKEN iTHES Project, MURI Center for Dynamic Magneto-Optics via the AFOSR Award No. FA9550-14-1-0040, the Japan Society for the Promotion of Science (KAKENHI), the IMPACT program of JST, CREST grant No. JPMJCR1676, the John Templeton Foundation, and the Australian Research Council.
\end{acknowledgments}

\end{document}